\documentclass[12pt]{article}
\newcommand{\interlinia}{}
\newcommand{\url}[1]{\texttt{#1}}

\title{Python for education: the exact cover problem}

\author{A. Kapanowski\\
{\em Marian Smoluchowski Institute of Physics,}\\
{\em Jagellonian University, ulica Reymonta 4,}\\
{\em 30-059 Krak\'{o}w, Poland}  }

\begin{document}
\maketitle

\begin{abstract}
\interlinia
Python implementation of Algorithm X by Knuth is presented.
Algorithm X finds all solutions to the exact cover problem.
The exemplary results for pentominoes, Latin squares and Sudoku
are given.
\end{abstract}

\interlinia

\section{Introduction}
\label{sec:intro}

Python is a powerful dynamic programming language that is used 
in a wide variety of application domains
\cite{python}.
Its high level data structures and clear syntax make it an ideal
first programming language
\cite{thinkpython}
or a language for easy gluing
together tools from different domains to solve complex problems
\cite{[2006_Langtangen]}.
The Python standard library and third party modules can speed up
programs development and that is why Python is used
in thousands of real-world business applications around the world,
Google and YouTube, for instance.
The Python implementation is under an opes source licence that make 
it freely usable and distributable, even for commercial use.

Python is a useful language for teaching even if students have
no previous experience with it.
They can explore complete documentation, both integrated into 
the language and as separate web pages.
Since Python is interpreted, students can learn the language
by executing and analysing individual commands.
Python is sometimes called "working pseudocode" because it is possible
to explain an algorithm by means of Python code and next
to run a program in order to check if it is correct.
Our aim is to use Python to implement an algorithm of solving
the exact cover problem. We prove that Python code is readable
and can be used to solve many medium size problems
in reasonable time.

The paper is organized as follows.
In Section~2 the exact cover problem is defined.
In Section~3 Python implementation of Algorithm X is presented.
Sections 4, 5, and 6 are devoted pentominoes, Latin squares,
and Sudoku, respectively.
A summary and conclusions are contained in Section~7.

\section{The exact cover problem}

In mathematics, given a collection S of subsets of a set X, 
an exact cover is a subcollection S* of S such that each element 
in X is contained in exactly one subset in S*.
In computer science, the exact cover problem is a decision problem 
to find an exact cover or else determine none exists. 
The exact cover problem is NP-complete
\cite{wiki_exact_cover}.

The relation "contains" can be represented by an incidence matrix $A$.
The matrix includes one row for each subset in S and one column 
for each element in X. The entry in a particular row and column is 1 
if the corresponding subset contains the corresponding element, 
and is 0 otherwise.
In the matrix representation, an exact cover is a selection of rows 
such that each column contains a 1 in exactly one selected row.

Interesting examples of exact cover problems are:
finding Pentomino tilings,
finding Latin squares, and
solving Sudoku.

The standard exact cover problem can be generalized to involve 
not only "exactly-one" constraints but also "at-most-one" constraints.
The N queens problem is an example of such generalization.

\section{Python implementation of Algorithm X}

Algorithm X is a recursive, nondeterministic,
backtracking algorithm (depth-first search)
that finds all solutions to the exact cover problem. 
Knuth efficiently implemented his Algorithm X by means of 
the technique called Dancing Links (DLX)
\cite{[2000_Knuth]}.
Algorithm X functions as follows.

\begin{verse}
If the matrix $A$ is empty, the problem is solved; terminate successfully. \\
Otherwise choose a column $c$ (deterministically). \\
Choose a row $r$ such that $A[r,c] = 1$ (nondeterministically). \\
Include row $r$ in the partial solution. \\
For each column $j$ such that $A[r, j] = 1$, \\
\hspace{0.5in}
delete column $j$ from matrix $A$; \\
\hspace{0.5in}
for each row $i$ such that $A[i, j] = 1$, \\
\hspace{1.in}
delete row $i$ from matrix $A$. \\
Repeat this algorithm recursively on the reduced matrix $A$.
\end{verse}

Now we would like to present Python implementation of the Algorithm X.
Extensive use of list comprehensions is present.
The program was tested under Python 2.5.
Let us define the exception \verb|CoverError| and 
the function to read the incident matrix
from a text file to the table $A$.
The table $A$ is represented by the list of nodes,
where a node is a pair (row, column) for a 1 in the incident matrix.
Any line of the text file should contain labels of incident matrix columns
with 1 in a given row.

\begin{verbatim}
class CoverError(Exception):
    """Error in cover program."""
    pass
def read_table(filename):
    """Read the incident matrix from a file."""
    f = open(filename,"r")
    table = []
    row = 0
    for line in f:
        row = row + 1
        for col in line.split():
            table.append((row, col))
    f.close()
    return table
A = read_table("start.dat")
\end{verbatim}
Let us define some useful global variables: 
\verb|B| to keep the solution (selected rows of the incident matrix),
\verb|updates| to count deleted nodes on each level,
\verb|covered_cols| to remember if a given column is covered.
The number of removed nodes is proportional to the number
of elapsed seconds. The 2 GHz Intel Centrino Duo laptop
did from 20 to 40 kilo-updates per second.
\begin{verbatim}
B = {}
updates = {}
covered_cols = {}
for (r, c) in A: covered_cols[c] = False
\end{verbatim}
Some functions to print the solution and to choose the next
uncovered column. In our program a column with the minimal
number of rows is returned because it leads to the fewest branches.
\begin{verbatim}
def print_solution():
    """Print the solution - selected rows."""
    print "SOLUTION", updates
    for k in B:
        for node in B[k]:
            print node[1],
        print
def choose_col():
    """Return an uncovered column with the minimal number of rows."""
    cols = [c for c in covered_cols if not covered_cols[c]]
    if not cols:
        raise CoverError("all columns are covered")
    # Some columns can have no rows.
    tmp = dict([(c,0) for c in cols])
    for (r,c) in A:
        if c in cols:
            tmp[c] = tmp[c] + 1
    min_c = cols[0]
    for c in cols:
        if tmp[c] < tmp[min_c]:
            min_c = c
    return min_c
\end{verbatim}
The most important is a recursive function \verb|search(k)| which
is invoked initially with $k=0$.
\begin{verbatim}
def search(k):
    """Search the next row k in the table A."""
    if not A:   # A is empty
        for c in covered_cols:
            if not covered_cols[c]:   # blind alley
                return
        print_solution()
        return
    c = choose_col()
    # Choose rows such that A[r,c]=1.
    rows = [node[0] for node in A if node[1]==c]
    if not rows:   # blind alley
        return
    for r in rows:
        box = []         # a place for temporaly removed rows
        # Include r in the partial solution.
        B[k] = [node for node in A if node[0]==r]
        # Remove row r from A.
        for node in B[k]:
            box.append(node)
            A.remove(node)
            updates[k] = updates.get(k,0) + 1
        # Choose columns j such that A[r,j]==1 (c is included).
        cols = [node[1] for node in B[k]]
        for j in cols:
            covered_cols[j] = True
            # Choose rows i such that A[i,j]==1.
            rows2 = [node[0] for node in A if node[1]==j]
            # Remove rows i from A to box.
            tmp = [node for node in A if node[0] in rows2]
            for node in tmp:
                box.append(node)
                A.remove(node)
                updates[k] = updates.get(k,0) + 1
        search(k+1)
        # Restore deleted rows.
        for node in box:
            A.append(node)
        del box
        del B[k]
        # Uncover columns.
        for j in cols:
            covered_cols[j] = False
    return
\end{verbatim}

The program can be saved to the file {\em cover.py }.
Next sections are devoted to the selected applications of the program.

\section{Pentomino}

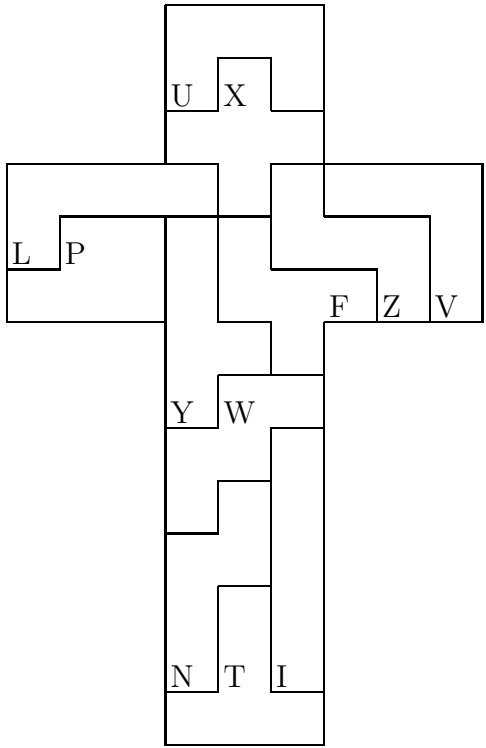
\begin{figure}
\begin{center}
\setlength{\unitlength}{20pt}
\begin{picture}(9,14)
%\put(3,0){\line(1,0){3}}
% obwiednia
\put(3,0){\line(1,0){3}}
\put(3,0){\line(0,1){10}}
\put(6,0){\line(0,1){8}}
\put(0,8){\line(1,0){3}}
\put(0,8){\line(0,1){3}}
\put(0,11){\line(1,0){4}}
\put(3,11){\line(0,1){3}}
\put(3,14){\line(1,0){3}}
\put(6,10){\line(0,1){4}}
\put(5,11){\line(1,0){4}}
\put(9,8){\line(0,1){3}}
\put(6,8){\line(1,0){3}}
% lewe ramie
\put(0,9){\line(1,0){1}}
\put(1,9){\line(0,1){1}}
\put(1,10){\line(1,0){4}}
% gora
\put(3,12){\line(1,0){1}}
\put(4,12){\line(0,1){1}}
\put(4,13){\line(1,0){1}}
\put(5,12){\line(0,1){1}}
\put(5,12){\line(1,0){1}}
\put(4,8){\line(0,1){3}}
\put(5,9){\line(0,1){2}}
% prawe
\put(5,9){\line(1,0){2}}
\put(7,8){\line(0,1){1}}
\put(8,8){\line(0,1){2}}
\put(6,10){\line(1,0){2}}
% dol
\put(4,8){\line(1,0){1}}
\put(5,7){\line(0,1){1}}
\put(4,7){\line(1,0){2}}
\put(4,6){\line(0,1){1}}
\put(3,6){\line(1,0){1}}
\put(3,4){\line(1,0){1}}
\put(4,4){\line(0,1){1}}
\put(4,5){\line(1,0){1}}
\put(5,6){\line(1,0){1}}
\put(4,3){\line(1,0){1}}
\put(3,1){\line(1,0){1}}
\put(5,1){\line(1,0){1}}
\put(4,1){\line(0,1){2}}
\put(5,1){\line(0,1){5}}
%\put(3.1,0.1){T}
\put(4.1,1.1){T}
\put(3.1,1.1){N}
\put(5.1,1.1){I}
%\put(3.1,4.1){W}
\put(4.1,6.1){W}
\put(3.1,6.1){Y}
%\put(5.1,7.1){F}
\put(6.1,8.1){F}
%\put(0.1,8.1){P}
\put(1.1,9.1){P}
\put(0.1,9.1){L}
%\put(4.1,10.1){X}
\put(4.1,12.1){X}
\put(3.1,12.1){U}
\put(7.1,8.1){Z}
\put(8.1,8.1){V}
\end{picture}
\end{center}
\caption[The 12 pentominoes form a cross.]{
\label{cross}
\interlinia
The 12 pentominoes form a cross.
There are 21 unique solutions.
The naming convention is also shown.}
\end{figure}

Polyominoes are shapes made by connecting certain numbers 
of equal-sized squares, each joined together with at least
one other square along an edge
\cite{[1994_Golomb]}.
Pentominoes are made from five squares and they can form
twelve distinctive patterns. 
Some letter names are recommended for them according to the shapes.
All pentominoes can fill a board
with 60 squares and of different shapes.
The standard boards are rectangles of $6 \times 10$, $5 \times 12$, 
$4 \times 15$, and $3 \times 20$,
but we can try a cross or a chessboard without 
the center four squares, see Fig. \ref{cross}.
Pentominoes can be rotated (turned 90, 180,
or 270 degrees) or reflected (flipped over).
Note that one-side pentominoes can be also considered,
where the reflection in forbidden.

The problem of forming a 60 square board with twelve pentominos
involves two kinds of constraints:

\begin{description}
\item[Pentomino]
For each of the 12 pentominoes, it must be placed exactly once. 
Columns names correspond to the pentominos: 
F, I, L, P, N, T, U, V, W, X, Y, Z.
\item[Square]
For each of the 60 squares, it must be covered by a pentomino exactly once. 
A square name can be its successive number.
\end{description}
Thus there are $12+60 = 72$ constraints in all.
Our results are collected in Table \ref{tab:pento}.

\begin{table}
\caption[Results for different pentomino boards.]{
\label{tab:pento}
\interlinia
Results for different pentomino boards.
Input is the size of the incident matrix, Solutions are all
solutions found by the program, Unique are different solutions,
and Updates are numbers of temporaly removed nodes.
The Cross board is shown in Fig. \ref{cross}.
The Chess board is $8 \times 8$ without the center four squares.}
\begin{center}
\begin{tabular}{ccrrr}
\hline\hline
Board & Input & Solutions & Unique & Updates
\\ \hline
$3 \times 20$  & $1100 \times 72$  & 8  & 2  & 9,770,304  \\
$4 \times 15$  & $1558 \times 72$  & 1472  & 368  & 237,324,570  \\
$5 \times 12$  & $1806 \times 72$  & 4040  & 1010  & 682,909,158  \\
$6 \times 10$  & $1928 \times 72$  & 9356  &  2339 & 1,296,313,446  \\
Cross & $1413 \times 72$  & 42  & 21  & 15,806,634  \\
Chess  & $1568 \times 72$  & 520  & 65  & 127,145,172  \\
\hline\hline
\end{tabular}
\end{center}
\end{table}

There are many other problems connected with pentominoes
that can be solved by means of the cover program.
Some of them were collected by G. E. Martin in his book
\cite{[1996_Martin]}: the Double Duplication Problem,
the Triplication Problem, for instance.
%D. Knuth considered the problem of packing 45 Y pentominoes
%into $15 \times 15$ board. Here, there are 225 constraints
%connected with the squares and 212 solutions.

\section{Latin square}

Latin square is an $n \times n$ table filled with $n$ different symbols
(for example, numbers from 1 to $n$) in such a way that each symbol 
occurs exactly once in each row and exactly once in each column.
An exemplary Latin square $4 \times 4$ is shown in Fig. \ref{latin4x4}.
Latin squares are used in the design of experiments and 
error correcting codes
\cite{wiki_latin}.

The problem of finding Latin squares
involves three kinds of constraints:

\begin{description}
\item[Square]
Each square must contain exactly one number (column name ij).
\item[Row-Number]
Each row must contain each number exactly once (column name RxNy).
\item[Column-Number]
Each column must contain each number exactly once (column name CxNy).
\end{description}
There are $3 n^2$ constrains and
the incident matrix is $n^3 \times 3 n^2$.
The rows describing the Latin square shown in Fig. \ref{latin4x4} are

\begin{figure}
\begin{center}
\begin{tabular}{|cccc|}
\hline
1 & 2 & 3 & 4 \\
2 & 3 & 4 & 1 \\
3 & 4 & 1 & 2 \\
4 & 1 & 2 & 3 \\ \hline
\end{tabular}
\end{center}
\caption[Latin square $4 \times 4$.]{
\label{latin4x4}
\interlinia
Latin square $4 \times 4$ normalized.
There are 4 unique solutions.}
\end{figure}

\begin{verbatim}
11 R1N1 C1N1
12 R1N2 C2N2
13 R1N3 C3N3
14 R1N4 C4N4
21 R2N2 C1N2
22 R2N3 C2N3
23 R2N4 C3N4
24 R2N1 C4N1
31 R3N3 C1N3
32 R3N4 C2N4
33 R3N1 C3N1
34 R3N2 C4N2
41 R4N4 C1N4
42 R4N1 C2N1
43 R4N2 C3N2
44 R4N3 C4N3
\end{verbatim}

A Latin square is normalized if its fist row and first column
are in natural order. For each $n$, the number of all Latin squares
is $n!(n-1)!$ times the number of normalized Latin squares.
The exact values are known up to $n=11$
\cite{[2005_McKay]}.
Our results for normalized Latin squares
are collected in Table \ref{tab:latin}.

\begin{table}
\caption[Results for normalized Latin squares.]{
\label{tab:latin}
\interlinia
Results for normalized Latin squares.
Input is the size of the incident matrix, Solutions are all
solutions found by the program, 
and Updates are numbers of temporaly removed nodes.}
\begin{center}
\begin{tabular}{ccrr}
\hline\hline
Board & Input & Solutions  & Updates
\\ \hline
$1 \times 1$ & $1 \times 1$      & 1  & 1             \\
$2 \times 2$ & $5 \times 12$     & 1  & 12            \\
$3 \times 3$ & $17 \times 27$    & 1  & 33           \\
$4 \times 4$ & $43 \times 48$    & 4  & 216          \\
$5 \times 5$ & $89 \times 75$    & 56  & 3,909            \\
$6 \times 6$ & $161 \times 108$  & 9,408  & 675,513         \\
$7 \times 7$ & $265 \times 147$  & 16,942,080  & 1,307,277,285      \\
$8 \times 8$ & $407 \times 192$  & 535,281,401,856  & ?      \\
$9 \times 9$ & $593 \times 243$  & 377,597,570,964,258,816  & ?      \\
\hline\hline
\end{tabular}
\end{center}
\end{table}

\section{Sudoku}

A standard Sudoku is like an order-9 Latin square,
differing only in its added requirement that each subgrid (box)
contain the numbers 1 through 9
\cite{[2006_Delahaye]}.
Generaly, a Sudoku of order $k$ ($n=k^2$) is an $n \times n$ table
which is subdivided into $n$ $k \times k$ boxes.
Each raw, column, and box must contain each of the numbers
1 through $n$ exactly once.
Any valid Sudoku is a valid Latin square.
An exemplary Sudoku $4 \times 4$ is shown in Fig. \ref{sudoku4x4}.
Note that the Latin square shown in Fig. \ref{latin4x4}
is not a valid Sudoku.

A Sudoku delivers many interesting and sometimes difficult
logic-based problems.
Let us start from the problem of counting the number of
valid Sudoku tables.
The problem involves four kinds of constraints:

\begin{description}
\item[Square]
Each square must contain exactly one number (column name ij).
\item[Row-Number]
Each row must contain each number exactly once (column name RxNy).
\item[Column-Number]
Each column must contain each number exactly once (column name RxNy).
\item[Box-Number]
Each box must contain each number exactly once (column name BxNy).
\end{description}
For the sudoku board $n \times n$, there are $4 n^2$ constrains and
the incident matrix is $n^3 \times 4 n^2$.
The exemplary rows for the Sudoku shown in Fig. \ref{sudoku4x4} are

\begin{figure}
\begin{center}
\begin{tabular}{|cc|cc|}
\hline
1 & 2 & 3 & 4 \\
3 & 4 & 1 & 2 \\ \hline
2 & 1 & 4 & 3 \\ 
4 & 3 & 2 & 1 \\ \hline
\end{tabular}
\end{center}
\caption[Sudoku $4 \times 4$.]{
\label{sudoku4x4}
\interlinia
Sudoku $4 \times 4$. There are 288 unique solutions.}
\end{figure}

\begin{verbatim}
11 R1N1 C1N1 B1N1
12 R1N2 C2N2 B1N2
13 R1N3 C3N3 B2N3
14 R1N4 C4N4 B2N4
21 R2N3 C1N3 B1N3
22 R2N4 C2N4 B1N4
23 R2N1 C3N1 B2N1
24 R2N2 C4N2 B2N2
31 R3N2 C1N2 B3N2
32 R3N1 C2N1 B3N1
33 R3N4 C3N4 B4N4
34 R3N3 C4N3 B4N3
41 R4N4 C1N4 B3N4
42 R4N3 C2N3 B3N3
43 R4N2 C3N2 B4N2
44 R4N1 C4N1 B4N1
\end{verbatim}

Our results are collected in Table \ref{tab:sudoku}.
A detailed calculation of the number of classic $9 \times 9$ Sudoku
solution was provided by Felgenhauer and Jarvis in 2005
\cite{[2005_Felgenhauer_Jarvis]}. 
The number is 6,670,903,752,021,072,936,960, or approximately
$6.67 \times 10^{21}$. This is $1.2 \times 10^{-6}$ times the number
of $9 \times 9$ Latin squares.
Felgenhauer and Jarvis identified 44 classes of different solutions,
where first three rows are fixed for a given class
when we are looking for solutions.

\begin{table}
\caption[Results for Sudoku.]{
\label{tab:sudoku}
\interlinia
Results for Sudoku.
Input is the size of the incident matrix, Solutions are all
solutions found by the program, 
and Updates are numbers of temporaly removed nodes.
Solutions for the $9 \times 9$ board are cited from the paper by
Felgenhauer and Jarvis
\cite{[2005_Felgenhauer_Jarvis]}.}
\begin{center}
\begin{tabular}{ccrr}
\hline\hline
Board & Input & Solutions  & Updates
\\ \hline
$1 \times 1$ & $1 \times 1$      & 1  & 1             \\
$4 \times 4$ & $64 \times 64$     & 288  & 21,712     \\
$9 \times 9$ & $729 \times 324$    & $6.7 \times 10^{21}$  & ?           \\
$16 \times 16$ & $4096 \times 1024$    & ?  & ?          \\
\hline\hline
\end{tabular}
\end{center}
\end{table}

A Sudoku puzzle is a partially completed table, which has a unique
solution and has to be completed by a player.
The problem of the fewest givens that render a unique solution
is unsolved, although the lowest number yet found is 17.
There are collected more than 38,000 17-Clou puzzles
and there is one known 16-Clue puzzle with two solutions.
Our program can easily complete a puzzle or can check that the unique
solution exists in few seconds.
Many puzzle enthusiasts are looking for the hardest Sudoku, i.e.
the  Sudoku which is the most difficult to solve for some solver programs.
The hardest Sudoku for our program was 21-Clue Sudoku called col-02-08-071
\cite{wiki_sudoku}
shown in Fig. \ref{hardest}.

\begin{figure}
\begin{center}
\begin{tabular}{|ccc|ccc|ccc|}
\hline
. & 2 & . & 4 & . & 3 & 7 & . & . \\
. & . & . & . & . & . & . & 3 & 2 \\
. & . & . & . & . & . & . & . & 4 \\ \hline
. & 4 & . & 2 & . & . & . & 7 & . \\
8 & . & . & . & 5 & . & . & . & . \\
. & . & . & . & . & 1 & . & . & . \\ \hline
5 & . & . & . & . & . & 9 & . & . \\
. & 3 & . & 9 & . & . & . & . & 7 \\
. & . & 1 & . & . & 8 & 6 & . & . \\ \hline
\end{tabular}
\end{center}
\caption[The hardest Sudoku $9 \times 9$.]{
\label{hardest}
\interlinia
The hardest Sudoku $9 \times 9$.
There are 113,072 updates in our program.}
\end{figure}

\section{Conclusions}
\label{sec:conclusions}

In this paper, we presented Python implementation of Algorithm X
solving the exact cover problem.
It has less than one hundred lines, counting comments.
The program can be used to solve any medium size problem
that can be formulated as the exact cover problem.
It can handle the cases without solutions or with multiple
solutions.

The program was used to solve some puzzles, to generate Latin
squares or Sudoku boards. The problems can be analysed according
to different criteria: the incident matrix size, number of 1
in a row, number of solutions, or a number of updates
on any level of backtracking.

The presented implementation of Algorithm X can be easily
extended to the case of "at-most-one" constraints.
We hope that the presented program will be used for teaching
or just for fun.

%\listoffigures

%\newpage

%\listoftables

%\newpage

\end{document}